\documentclass[twocolumn,prb]{revtex4}

\usepackage{amsmath}
\usepackage{graphicx}
\usepackage{epsfig}

\def\e0{{e_0}}

 \def\bea{\begin{eqnarray}}
 \def\eea{\end{eqnarray}}
\def\beq{\begin{equation}}
\def\eeq{\end{equation}}

\newcommand{\ind}{\hspace*{\parindent}}

\begin{document}
\title{Corked Bats, Juiced Balls, and Humidors:\\  The Physics of Cheating in Baseball}

\author{Alan M. Nathan}
\affiliation{Department of Physics, University of Illinois,
Urbana, Illinois 61801} \email{a-nathan@uiuc.edu}

\author{Lloyd V. Smith and Warren L. Faber} \affiliation{School of Mechanical and
Materials Engineering, Washington State University, Pullman,
Washington 99164} \email{lvsmith@mme.wsu.edu}

\author{Daniel A. Russell} \affiliation{Department of Physics, Kettering University, Flint, Michigan 48504}
\email{drussell@kettering.edu}

\date{\today}

\begin{abstract}

Three separate questions of relevance to Major League Baseball are
investigated from a physics perspective. First, can a baseball be
hit farther with a corked bat?  Second, is there evidence that the baseball is
more lively today than in earlier years?  Third, can
storing baseballs in a temperature- or humidity-controlled environment significantly affect
home run production?
Each of these questions is
subjected to a physics analysis, including an experiment, an
interpretation of the data, and a definitive answer.  The answers to the three questions
are no, no, and yes.

\end{abstract}

\maketitle

\section{Introduction}\label{sec:intro}
\ind Baseball is rich in phenomena that are ripe for a physics
analysis.  In the last decade this journal has seen an explosion
in the number of papers addressing interesting issues in baseball
from a physics perspective.  In this paper we address three new
issues of relevance to Major League Baseball (MLB).
  While the topics are seemingly separate, they all involve
the common physics issue of the ball-bat collision and all
are explored using variations of the same experimental
technique described in Sec.\ref{sec:SSL}.  We first
investigate whether or not a baseball can
be hit harder and therefore farther with an illegally modified
corked bat (Sec.~\ref{sec:cork}).  We next investigate whether there
is any direct evidence that the baseball of today is more or less
lively than the baseball of yesteryear (Sec.~\ref{sec:juice}). Finally
 we investigate whether a baseball stored at
elevated temperature or humidity will lead to fewer home runs (Sec.\ref{sec:humidor}).   We conclude
the paper with a brief summary in Sec.~\ref{sec:summary}.

\section{Description of the Bat and Ball Test Facility}
\label{sec:SSL} 
\begin{figure}[htb]
\begin{center}
\includegraphics[angle=0,width=0.4\textwidth]{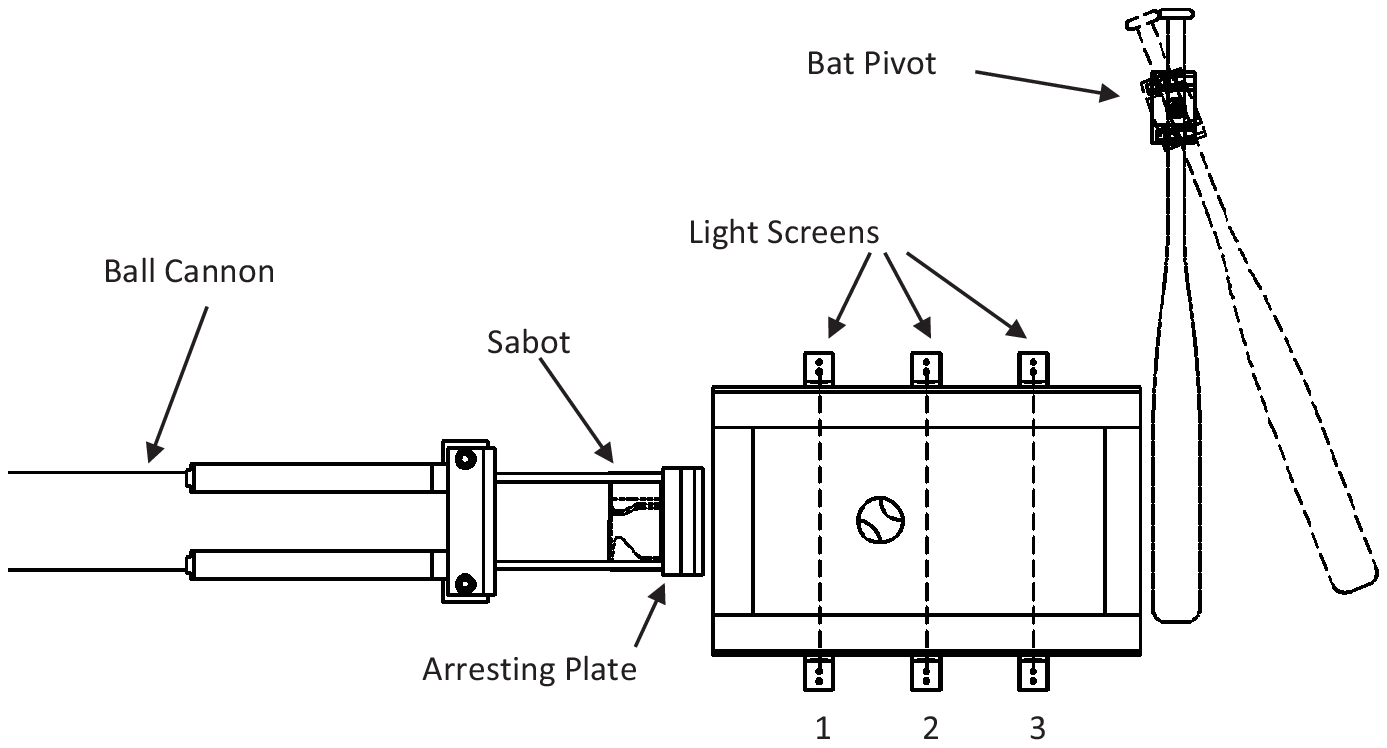}
\caption{\label{fig:cannon}Schematic of the bat and ball testing
facility at the Sport Science Laboratory with a bat as the impact
surface.  For some of the studies, the bat was replaced by a fixed
rigid surface, either flat or cylindrical.}
\end{center}
\end{figure}

\ind All the experimental work for these studies
were done at the bat and ball test facility at the Sports Science
Laboratory.\cite{smith08} The experimental setup is depicted
schematically in Fig.~\ref{fig:cannon}   The measurements
consisted of firing a baseball from a high-speed air
cannon onto a stationary impact surface.  While inside the barrel
of the cannon, the ball traveled in a sabot which allowed control
of the ball speed and orientation.  An arresting plate at the end of the cannon
captured the sabot while allowing the ball to continue unimpeded.
Three light screens were placed between the cannon and impact
surface to measure the speed of the incident ($v_0$) and
rebounding ($v_f$) ball. The location of the impact surface
relative to the cannon was adjusted so that the ball rebound path
was within 5$^\circ$ of the inbound path. The air pressure to the
cannon was adjusted to achieve an incident speed within
1~mph\cite{romer99} of the target speed.  The laboratory was maintained at fixed
72F temperature and 50\% relative humidity for all of the impact measurements.

Three different impact surfaces were used in the studies: a
baseball bat shown in Fig.~\ref{fig:cannon}, a fixed rigid flat
surface, and a fixed rigid cylindrical surface.  The bat was used
for the corked-bat studies and for some of the juiced-ball
studies. The bat was mounted horizontally and supported by
clamping it at the handle to a structure that was free to pivot
about a vertical axis located six inches from the knob.  The
collision efficiency $e_A$ is given by \beq e_A\, = \,
\frac{v_f}{v_0} \, = \, \frac{e-m/M_{eff}}{1+m/M_{eff}} \, ,
\label{eq:ea} \eeq where $e$ is the ball-bat coefficient of
resitution (BBCOR) and $m$ is the mass of the ball.\cite{nathan03}
The effective mass of the bat is $M_{eff}=I_6/z^2$, where $I_6$ is
the moment of inertia of the bat about the pivot, and $z$ is the
distance from the impact location to the pivot.   Eq.~\ref{eq:ea}
can be inverted to find the BBCOR value from measurements of $v_0$
and $v_f$.
%The bat was impacted six times at
%half-inch increments on the barrel to find the location of maximum
%BBCOR, where all subsequent impacts were done.

For the fixed rigid surfaces, $M_{eff}\rightarrow\infty$, so that
the COR is just the ratio of outgoing to incoming speed.  Under
these conditions, $e$ is referred to as the ball COR for the flat
surface or the ball cylindrical COR, or CCOR, for the cylindrical
surface. The flat surface was used for some of the juiced-ball
studies to measure the ball COR. The cylindrical surface was used
in the humidor studies for determining the ball CCOR, which is a
better approximation to the forces and deformation encountered in
a ball-bat collision.  The 2.63 inch diameter of the impact surface was
chosen to closely approximate the diameter of a baseball bat.  A
relatively slow 60 mph incident speed was used for the CCOR
measurements to minimize ball degradation during the study.
%Also
%measured was the dynamic stiffness of the ball, using an array of
%three piezoelectric load cells placed between the cylindrical
%surface and the rigid wall to measure the impact force.  The
%stiffness of the ball $k$ is related to the peak impact force $F$
%and the initial velocity by
% \begin{equation}
% k\,=\,\frac{1}{m}\left (\frac{F}{v_0}\right )^{1/2}
% \end{equation}
%Ball stiffness is of interest in amateur play where it affects the
%trampoline effect observed with hollow bats. The reported dynamic
%stiffness and CCOR are averages over six valid impacts from each
%ball.

\section{Can a Baseball be hit farther with a Corked Bat?}
\label{sec:cork}
\subsection{Introduction}

In early June during the 2003 Major League Baseball season, an
event occurred that dominated the sports news for several days.
Sammy Sosa, the Chicago Cubs slugger, was caught using an
illegally ``corked'' bat during a game.  This event offered a rare
opportunity for scientists to comment on events in the world of
sports by addressing the question of whether corking a bat
gives the batter an advantage.

A corked bat is a wood bat in which a cylindrical cavity is
drilled axially into the barrel of the bat.  Typically the
diameter and length of the cavity are approximately one and
ten inches, respectively. The cavity is then filled with a light inert material
such as cork (hence, corking), the goal being to disguise the fact
that the bat has been illegally modified.  By removing weight from
the barrel region, the batter can achieve a higher swing speed.
However, the lower barrel weight implies a lower collision
efficiency. Therefore, if the goal of the batter is to achieve as
high a batted-ball speed (BBS) as possible, the increased swing
speed is at least partially compensated by the less effective
collision. One goal of the present study is to investigate the
tradeoff between swing speed and collision efficiency to see
whether a batter can achieve a higher BBS with a corked bat.

Corking a bat may offer another advantage, the so-called ``trampoline effect,''\cite{tramp04}
at least according to anecdotal claims by some batters.  The
trampoline effect occurs in hollow metal bats due to the ability
of the thin wall of the bat to compress when in contact with the
ball, thereby increasing the elasticity of the collision.  The
increased elasticity results in a larger collision efficiency
and--all other things equal--a larger BBS.  The second goal of this study
is to determine whether such a trampoline effect exists for a hollow or corked wood bat.

\subsection{Experimental Procedures}

These issues were addressed at the ball-bat test facility
described in Sec.~\ref{sec:SSL}.   Impact measurements were
performed which consisted of firing a baseball from the air cannon
at a speed of approximately 110 mph onto a stationary bat. The
speed of the incoming and rebounding baseball ($v_0$ and $v_f$,
respectively) were measured, and Eq.~\ref{eq:ea} was used to determine
$e_A$ and the BBCOR.  For a given BBCOR
reducing the moment of inertia by corking the bat also reduces $M_{eff}$ and therefore $e_A$, as discussed
earlier.

The properties of the bats used in the study are given in
Table~I., with the weight, center-of-mass, and moment of inertia
measured using standard techniques.\cite{smith08}
\begin{table}[h]
\begin{center}
\caption{Properties of the Rawlings 34-inch bats used in the
studies and the results of the measurements. The ``hollow'' and
``corked'' bats are modifications of the ``unmodified'' bat.  The
``control'' bat is a similar but different bat altogether.  The
weight is in ounces; the center of mass CM is measured in inches
from the knob end of the bat; and the moment of inertia I$_6$ is
with respect to a point six inches from the knob and in units of
oz-in$^2$.  The dimensionless quantities collision efficiency
$e_A$, $m/M_{eff}$, and coefficient of restitution $e$ are
measured at an impact location 29 inches from the knob. Estimated
uncertainties in the least significant digit are given in parentheses.} \vspace{0.1in}
\begin{tabular}{|c|cccccc|}\hline
 bat& weight & CM & I$_6$  & $m/M_{eff}$ &$e_A$ & BBCOR\\ \hline
unmodified & 30.6 & 23.5 &11635 & 0.2269  &0.214(2) & 0.490(2)\\
hollow & 27.6 & 22.9 & 10044 &0.2628& 0.173(2)& 0.481(2)\\
corked & 28.6 & 23.2 & 10659 &0.2477& 0.193(2)
& 0.488(2) \\
control & 31.1 & 23.6 & 12106 & 0.2181 & 0.227(1) & 0.494(2)\\
 \hline
 \end{tabular}
 \end{center}
 \label{tab:bats}
 \end{table}
The unmodified bat had a length of 34 inches and an unmodified weight
of 30.6 oz. First, the unmodified bat was impacted a total of six
times. Then a cavity one inch in diameter and ten inches deep was
drilled into the barrel of the bat, reducing the weight to 27.6
oz.  This ``hollow'' bat was impacted a total of six times.  Then
the cavity was filled with crushed-up pieces of cork, raising the
weight to 28.6 oz. The corked bat was impacted twelve times.  Then
the cork was removed and the drilled bat was impacted again five
times. Unfortunately, the bat broke at the handle on the last
impact.  We had intended to fill the cavity with superball
material, but that part of the experiment was cut short by
breaking the bat.  All impacts used the same baseball and all were
at the same location, five inches from the barrel end of the bat.
A twin ``control'' bat, with properties nearly identical to those
of the
unmodified bat, was impacted at various times throughout the
measurement cycle to verify that the properties of the ball did
not change in the course of the measurements.

\subsection{Results and Discussion}

\begin{figure}[htb]
%\begin{center}
\vspace{0.3in}
\epsfig{angle=0,width=0.4\textwidth,file=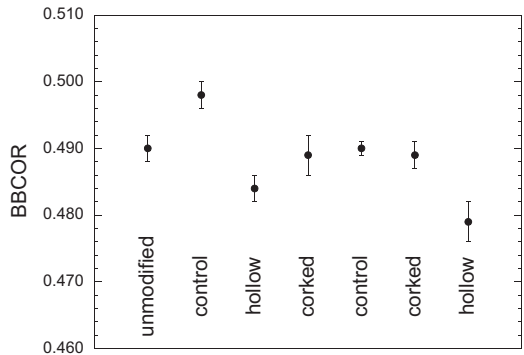}
\caption{\label{fig:bbcor}Results of the BBCOR measurements for
the corked bat study. Each point is the average of 5 or 6 impacts
and the ordering of the points is the same as the sequencing of
the measurements.}
%\end{center}
\end{figure}

\begin{figure}[htb]
%\begin{center}
\epsfig{angle=0,width=0.4\textwidth,file=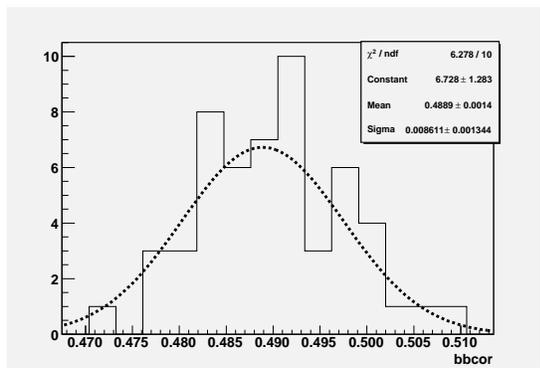}
\caption{\label{fig:cor-hist}Histogram of BBCOR measurements for
all the bats listed in Table I.  The dotted curve is a Gaussian
fit to the distribution, with a mean of 0.489 and a standard
deviation of 0.009.}
%\end{center}
\end{figure}

The results from each part of the measurement cycle are shown in
Fig.~\ref{fig:bbcor}.  Average values of both $e_A$ and BBCOR for
each bat are given in Table~I.  A histogram of the measured BBCOR
for 54 total impacts, including all bats, is shown in
Fig.~\ref{fig:cor-hist}, where we see that the mean BBCOR is 0.489
and the standard deviation is 0.009.   From these results, we
conclude that to within about 0.01 (or about 2\%), the BBCOR is
identical for all four bats listed in Table~I, despite the fact
that the spread in $e_A$ values is considerably larger, of order
0.05 (or over 20\%).  The BBCOR values for the corked and
unmodified bats are even closer in value and are statistically
indistinguishable to better than 0.6\%.  By comparison, the BBCOR
of a typical hollow aluminum bat exceeds that of a wood bat of
comparable dimensions by at least 10\%.  We conclude that there is
no evidence for a trampoline effect in a corked bat.

We next investigate the tradeoff between higher swing speed and lower
collision efficiency, utilizing the formalism described by
Nathan.\cite{nathan03} Accordingly, the batted-ball speed may be found using
 \beq BBS\, = \,
e_Av_{pitch}\,+\,(1+e_A)v_{bat} \, , \label {eq:bbs} \eeq which
relates the BBS to the pitch speed, the bat speed, and the
collision efficiency.  To compare a corked and unmodified bat, we
use the measured values of the collision efficiencies (Table~I),
along with a typical pitch speed
$v_{pitch}$=90 mph and the prescription for
bat speed\cite{nathan03} \beq v_{bat}\,=\,70\, {\rm mph}\left
(\frac{I_0}{I_{knob}}\right )^{n} \, , \label{eq:vbat} \eeq
where $I_{knob}$ is the moment of inertia of the bat about the
knob and $I_0$ is a reference moment of inertia.  The value of
$I_{knob}$ is determined from $I_6$, the location of the center of
mass (see Table I), and the parallel axis theorem.  We take
$I_0$ to be the moment of inertia about the knob of the
unmodified bat (19213 oz-in$^2$).  The rationale behind Eq.~\ref{eq:vbat} comes from
the observation\cite{greenwald01,fleisig02,koenig,smith03} that
the rotation axis of the bat just prior to meeting the ball is
about a point very close to the knob, so it is natural to expect
the bat speed to depend on $I_{knob}$.  The exponent $n$, which
characterizes how the bat speed depends on $I_{knob}$, is not
known from any first principles. However, as discussed by
Adair,\cite{adair02} one can confidently set two extreme limits
for $n$.  A lower limit $n$ = 0 is realized when the batter swings the
bat at the same speed, independent of $I_{knob}$; the limit $n$ =
0.5
is realized when the kinetic energy imparted to the bat is
independent of $I_{knob}$.  Experimental data from
baseball\cite{greenwald01,fleisig02,koenig} and slow-pitch
softball\cite{smith03} seem to be consistent with $n$ $\approx$ 0.25,
or halfway between the extreme limits.

\begin{figure}[htb]
%\begin{center}
%\vspace{0.4in}
\includegraphics[angle=0,width=0.4\textwidth]{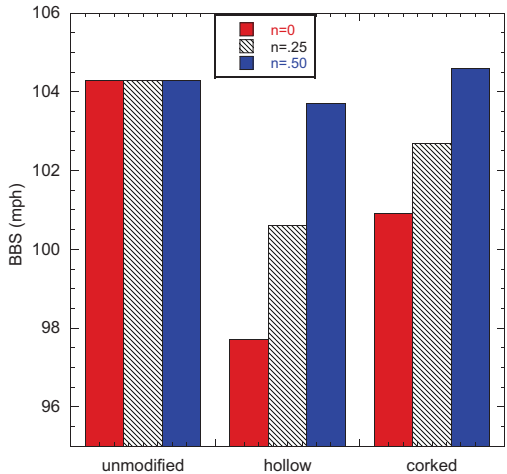}
\caption{\label{fig:bbs}Calculated value of batted ball speed
(BBS) for three different values of the scaling exponent $n$.}
%\end{center}
\end{figure}

Fig.~\ref{fig:bbs} shows the computed BBS for the unmodified,
hollow, and corked bats for $n$ = 0, 0.25, and 0.50.  Note that since
the moment of inertia of the unmodified bat was taken as $I_0$,
the BBS for that bat is independent of $n$.  This figure shows that
for all but the most extreme value $n$ $\approx$ 0.5, the BBS of the
unmodified bat {\it exceeds} that of the hollow or corked bat.  We
conclude that there is no advantage to corking a bat if the goal
is for the BBS to be as large as possible, as is the case for a
home run hitter.  Said a bit differently, a baseball cannot be hit
farther by corking a bat.  Indeed, one actually draws the opposite
conclusion, namely that corking almost always results in a lower
BBS and therefore a shorter fly ball distance.

It is worthwhile pointing out, however, that there are other
reasons why a batter might choose to cork a bat.  The smaller
moment of inertia results not only in a higher bat speed but most
likely in a higher bat acceleration; that is, in the parlance of
baseball, the batter can ``get around quicker.''  If a batter is
not a home run hitter but mainly a contact hitter, getting around
quicker offers a distinct advantage, since the batter can wait
longer on the pitch as well as more easily adjust the swing after
the swing has already begun. So, while corking may not allow a
batter to hit the ball farther, it may well allow a batter to hit
the ball solidly more often.  Indeed, while the present study shows 
that corked bats do not result in longer home runs, it
makes no statement about whether home runs might be hit more often
with a corked bat.

\section{Is the baseball juiced?}
\label{sec:juice}
\subsection{Introduction}
Is today's baseball juiced?  That is, is the baseball used in the
game today more lively than the baseball of yesteryear?  Or in
more scientific language, is the coefficient of restitution (COR)
of today's baseballs larger than that of earlier eras?  Regardless
of which version of the question is posed, the underlying issue is
one that periodically arises in the game of baseball, usually
during periods when there is a marked increase in the rate of home
run production.  The rules of MLB do not seem to be very
discriminating regarding the COR.  Those rules specify
that the COR of official baseballs must
lie in the range 0.514--0.578 when the ball is impacted on a flat plate at an
incident speed of 58 mph.  That range of acceptable values of $\pm
6\%$ leads to significant difference in performance when
extrapolated to the higher speed that are relevant to the game.
Our own estimate is that such a range would amount to a spread of
approximately 35 ft on the distance of a long fly, in agreement
with an earlier estimate by Kagan.\cite{kagan90}

Most recently the issue of juiced baseballs attracted widespread
attention during the early part of the 2000 MLB season.  During
April and May of that season, home runs were hit at a rate
markedly higher than the rate over the same time period in the
previous year. There was much speculation in the
baseball-observing community that the increase was due to the
juicing of the ball. As a result, MLB commissioned a study by the
Baseball Research Center at University of Massachusetts at Lowell
to compare the COR of baseballs from years 1998, 1999, and 2000.
Although not published in the refereed literature, the report was
widely disseminated.\cite{UML2000} Interestingly, the measurements done
at 58 mph were clustered at the upper end of the range allowed
by MLB.
 The principal conclusion of the
study was that there were no significant performance difference
among the three sets of baseballs. A subsequent study commissioned
by the Cleveland Plain Dealer (CPD) reached the same
conclusion.\cite{CPD2000}  However, the CPD study reported that
the COR of present-day baseballs is significantly larger than the
COR measured at the National Bureau of Standards in
1945.\cite{briggs}  This conclusion was based on a comparison of
the COR of 2000 balls at 89 mph (0.54, according to their
measurements) to the value of 0.46 for 1938 balls from the NBS
study, which was done at a slightly higher speed of 104
mph.\cite{smith10}

There are only two systematic studies of baseball COR of which we
are aware in the scientific literature.  First is that of Hendee,
{\it et al.}\cite{hendee} who measured the COR of various
baseballs up to speeds of 90 mph.  Second is that of Chauvin and
Carlson,\cite{chauvin} who measured up to 150 mph.  The results of
both studies are quite interesting in that they show that
baseballs with nearly identical COR's at the MLB-specified testing
speed of 58 mph can have considerably different COR's at higher
speeds. This result suggests that COR measurements at 58 mph might
have little relevance for the relative performance of baseballs at
the higher speeds at which the game is played.

The purpose of the present study is to compare the COR of
baseballs from different eras at as high a speed as practical.
The main difficulty in any measurement of this type is finding a
supply of unused baseballs from earlier years.  Serendipitously we
were able to find unopened boxes of baseballs from the late
1970's.  These baseballs were provided to us by the family of
Charlie Finley, then-owner of the Oakland A's, and were official
American League baseballs bearing the facsimile signature of
then-AL President Lee MacPhail and manufactured by Rawlings. These
facts constrain the baseballs to the period
1976-1980.\cite{baseballs}  The present-day baseballs were
purchased directly from Rawlings in 2004, the year the
measurements were actually performed.

\subsection{Experimental Procedures}
In the present study, the ball-bat test facility at the Sports Science Laboratory was used to fire
baseballs at speeds in the range 60-125 mph onto either a massive and
rigid flat steel plate or onto a wood bat.  For the massive plate,
the COR is just the ratio of rebound to
incident speed, both of which were measured.  For the bat, the same techniques
described in the corked bat study were used to extract the BBCOR.
It was assumed that the COR is essentially identical to the peak BBCOR,
where energy losses due to bat vibrations are minimal. The peak location was found
in a supplemental experiment using a different set of baseballs to
scan across the barrel of the bat.
All baseballs used in this study were conditioned by storing them in a
controlled 50\% relative humidity
environment for at least two weeks prior to the measurements.

\begin{figure}[htb]
\begin{center}
\epsfig{angle=0,width=0.4\textwidth,file=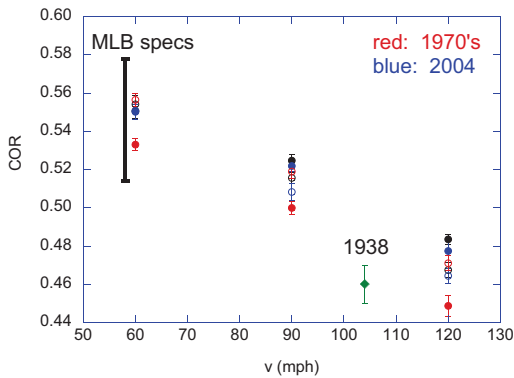}
\caption{Measured values of the ball COR vs incident speed for the
three 1970's balls (red) and the three
2004 balls (blue), with different plotting symbols corresponding to different balls. The vertical line at 58
mph is the range of COR specified by MLB. The
 diamond at 104 mph is the measurement of Briggs\cite{briggs} on
1938 baseballs.} \label{fig:corv}
\end{center}
\end{figure}

\begin{figure}[htb]
\begin{center}
\epsfig{angle=0,width=0.4\textwidth,file=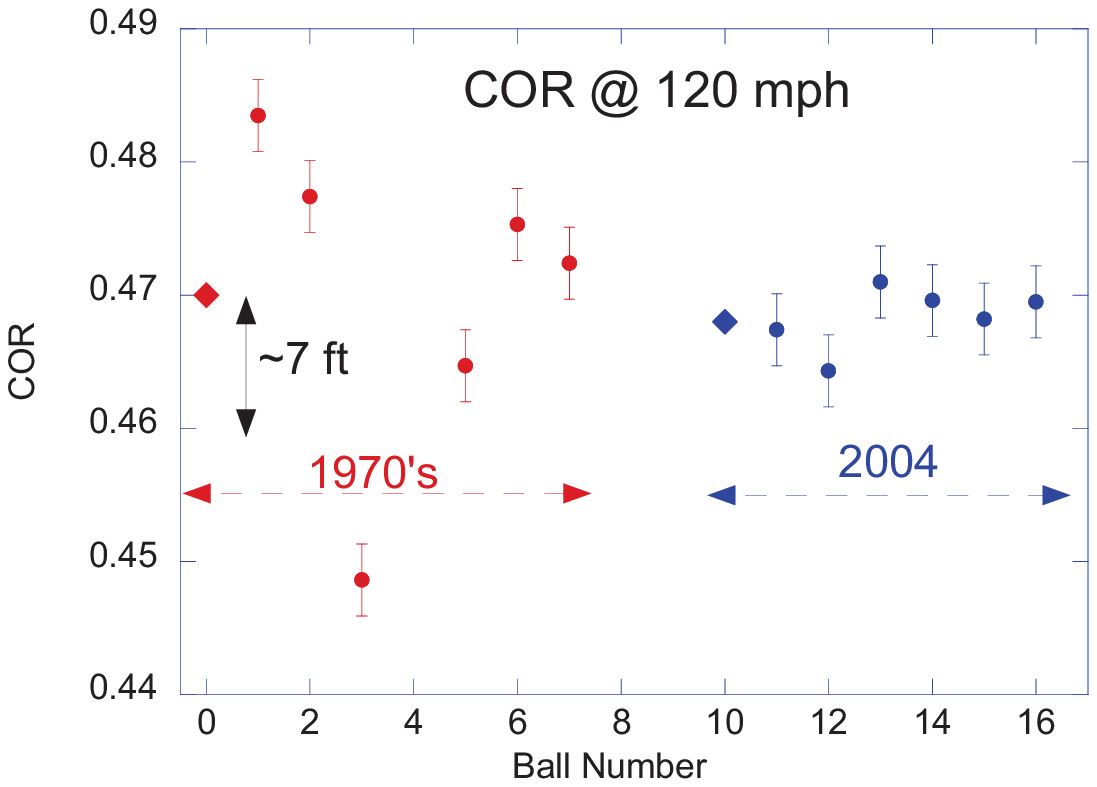}
\caption{Measured values of the COR of different balls at an
incident speed of 120 mph.  The red points (balls 1-3 and 5-7) are
from the 1970's and the blue points (balls 11-16) are from 2004.
The diamonds at 0 and 10 are averages over the six balls from each
era (0.470 and 0.468, respectively, for the 1970's and 2004
balls). The ``7 ft'' label is an estimate of the change in a long
fly ball distance due to a change in COR given by the length of
the double arrow (0.01).} \label{fig:cor120}
\end{center}
\end{figure}

\begin{figure}[htb]
\begin{center}
\epsfig{angle=0,width=0.4\textwidth,file=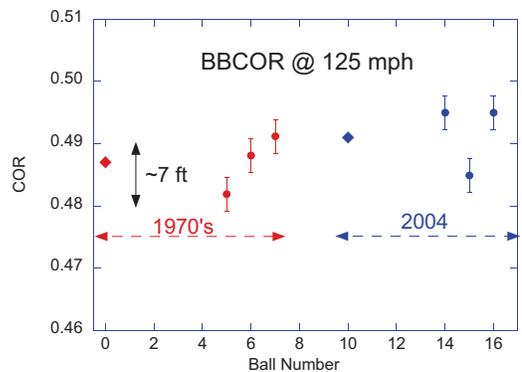}
\caption{Measured values of the COR of different balls at an
incident speed of 125 mph on a bat.  The red points (balls 5-7)
are from the 1970's and the blue points (balls 14-16) are from
2004.  The diamonds at 0 and 10 are averages over the three balls
from each era (0.487 and 0.491, respectively, for the 1970's and
2004 balls). The ``7 ft'' label is an estimate of the change in a
long fly ball distance due to a change in COR given by the length
of the double arrow (0.01).} \label{fig:corbat}
\end{center}
\end{figure}

The study was conducted in three parts.  First, the COR of three
baseballs from each set was measured by impacting the flat
plate at incident speeds of 60, 90, and 120 mph.  The results
are presented in Fig.~\ref{fig:corv}, where each point is an
average over four impacts.  Second, the COR of three additional
balls from each set were measured in the same manner at the fixed
incident speed of 120 mph, each ball being impacted four times.
The results are presented in Fig.~\ref{fig:cor120}, which also
includes the 120 mph results from the first part.  Third, the same
balls tested in the second part were tested again by impacting a
wood bat with a 125 mph initial speed.  The results of the ball-bat
impact study are presented in Fig.~\ref{fig:corbat}, where each
point again represents an average over three impacts.

\subsection{Results and Discussion}
We start this discussion with the primary conclusion evident from
an inspection of Fig.~\ref{fig:corv}-\ref{fig:corbat}.  There is
nothing in the current measurements to suggest any significant
difference in COR of the baseballs tested from the two different
eras. Indeed, averaging the 120-mph results from the ball-flat
plate collisions yields nearly identical results for the two sets
of balls (0.470 and 0.468).  Similarly, averaging the 125-mph
ball-bat collision results yields 0.487 and 0.491 for the older
and newer balls, respectively.

Some interesting secondary conclusions emerge from the data.
Referring to Fig.~\ref{fig:corv}, we see that all baseballs fall
within the rather broad range allowed by MLB at 60 mph.  Indeed,
with one exception, the COR of the balls tested at 60 mph fall within 0.01
of each other, a spread significantly smaller than the 0.064
range allowed by MLB.  The spread in values approximately doubles
at 120 mph, confirming the effect found by Hendee {\it et
al.}.\cite{hendee}.  The data also confirm the result from the CPD
study regarding the comparison with the NBS measurement (see
Fig.~\ref{fig:corv}.  If the present data are interpolated to find
an expected result at 104 mph, that result is larger than the
Briggs result by about 0.03.  One interesting observation from
Fig.~\ref{fig:cor120} is that the 2004 balls show much more
uniformity than the 1970's balls, perhaps pointing to better
quality control during the manufacturing process of today's
baseballs.

While our principal finding is the lack of evidence that today's
baseball is any more or less lively than that of an earlier time,
we caution the reader that this result applies only to the very
small sample of balls that we tested.  It is not possible to
extrapolate this result to make more general statements about the
relative liveliness of baseballs without more extensive testing.
Given the difficulty of obtaining older unused baseballs, we will
likely never be able to make such statements.

\section{What's the deal with the humidor?}
\label{sec:humidor}
\subsection{Introduction}
\ind Coors Field, home of the Colorado Rockies in mile-high
Denver, is well-known to be a pitcher's nightmare and a batter's
paradise. Because the air density in Denver is approximately 80\%
of that at sea level, fly balls carry farther and there is less
movement on pitches, both of which contribute to an increase in a
variety of offensive statistics.  However, starting in 2002 the
Colorado Rockies starting storing their baseballs in a humidor,
which was kept at a constant 70$^\circ$F temperature and 50\%
relative humidity (RH). Since that time, various offensive
statistics have dropped, such as home runs or total runs per
game.\cite{meyer08} So the question posed in the title to this
section arises. Rephrasing the question in a more scientific way,
is it plausible that the humidor could account for the decrease in
offensive statistics at Coors Field since 2002?

Two recent studies have addressed this question.  Meyer and
Bohn\cite{meyer08} investigated the effect of the humidor on the
aerodynamics of official MLB baseballs, specifically on the flight
of a fly ball.  Elevated humidity is expected to increase both the
weight and the diameter of a baseball. The authors investigated
these effects at RH values of 33\%, 56\%, and 75\%, then used
their findings along with models for the drag and lift to
calculate trajectories of batted and pitched baseballs.  For
batted balls, the expectation is that elevated humidity produces
two partially compensating effects.  An increased diameter results
in a larger drag force while an increased weight results in a
smaller drag acceleration. Indeed, their net result was an {\it
increase} of 2 ft in the distance travelled by a fly ball on a
typical home run trajectory when the RH is changed from 30\% to
50\%.  Not only is the effect too small to be significant, it even
goes in the wrong direction to account for the decrease in
offensive statistics. For pitched balls, the same two effects
result in slightly less movement on a pitched baseball for a given
velocity and spin. However, the authors speculated that the
ability of the pitcher to impart spin to the ball might be greatly
improved at the higher humidity, giving rise to more movement on
the pitch. Indeed, there is anecdotal information suggesting that
balls stored at very low RH tend to feel slippery to the pitcher,
making it difficult to put spin on the ball.

Meyer and Bohn recognized that a much larger effect on fly ball
distances comes from the change in the COR of the ball.  This
effect had been been investigated by Kagan and
Atkinson,\cite{kagan04} who measured the COR of an NCAA-approved
baseball at 61 mph over the range 0-100\% in RH.  They found that
the COR decreased by 0.054 over that range, from which they
estimated a decrease in fly ball distance by about 6 ft between
30\% and 50\% RH.

It is suspected that the COR increases with increasing temperature, leading to anecdotal reports
of equipment managers manipulating the COR of baseballs by using ``hotter'' balls when their team is batting and
``cooler'' balls when the opposing team is batting.  The only experimental data addressing this issue of which we are aware are those of Drane and Sherwood,\cite{drane04} who find that the flat-plate COR of a standard NCAA baseball measured at 60 mph increases from 0.524 to 0.548 when the temperature is increased from 25F to 120F.  To our knowledge, there are no comparable data for a MLB baseball.

The present experiment seeks to improve and extend the work of
Kagan and Atkinson with higher precision measurements of the CCOR of official MLB baseballs at more values of
the relative humidity.  In addition new data are presented
on the temperature
dependence of the CCOR.
%and ball stiffness.

\subsection{Experimental Procedures}

\begin{figure}[htb]
\begin{center}
\epsfig{angle=0,width=0.4\textwidth,file=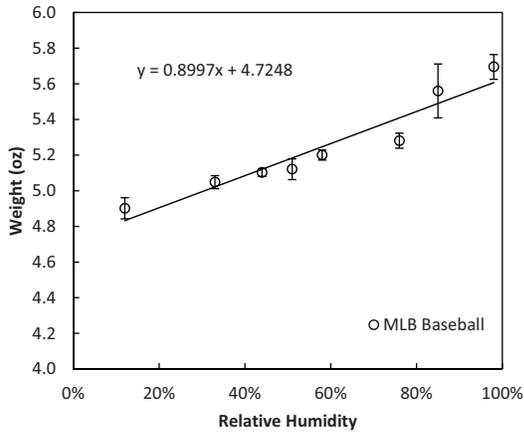} \caption{Ball
weight as function of relative humidity.}
 \label{fig:w}
\end{center}
\end{figure}

Eight groups of official MLB baseballs were given a controlled humidity exposure, after which their weight and
 CCOR were measured.
Four dozen baseballs were placed in a 50\% RH
conditioning environment for four months. The balls were then
divided into eight groups of six baseballs
and placed in separate conditioning chambers at relative
humidities ranging from 11\% to 97\% RH and at a fixed temperature of 72F for six weeks. The
humidity was controlled by suspending the balls over saturated
salt solutions in sealed containers. Humidity sensors were placed
in half of the containers to verify the target humidity level
 was maintained throughout the study. A control group was left in the initial 50\% RH
environment for the duration of the study.
To determine when the balls had reached saturation, the weight of
three baseballs (separate from the primary study) at
33\% and 97\% RH was continuously monitored.
%The weight change of
%a representative ball for each case is shown in Fig.~\ref{fig:lvs2}.
%Weight change is plotted vs. the square root of time to compare
%with Fickian diffusion.
The balls in the 33\% RH environment
reached saturation in just over a week, while the balls at 97\% RH
required upwards of three weeks. The average weight for each group
is shown in Fig.~\ref{fig:w} as a function of the humidity level.
The error bars represent the standard deviation for each group.
Meyer and Bohn reported a weight gain
of 3.4\% for MLB baseballs when the humidity was increased from
33\% to 75\%, in agreement with our measurement of 3.8\%.

To study the effect of temperature on the CCOR, the balls were conditioned at 50\%RH
and 72F for 4 months, then heated or cooled for 24 hours prior to testing.  The CCOR of the balls
were then quickly measured before any appreciable temperature change could occur.  Since the
oven and cooler used for the temperature tests were surrounded by air at 72F and 50\% RH
and the temperature excursion was relatively short, the moisture content of the balls was similar to their 50\% RH conditioned state.

\subsection{Results and Discussion}

\begin{figure}[htb]
\begin{center}
\epsfig{angle=0,width=0.4\textwidth,file=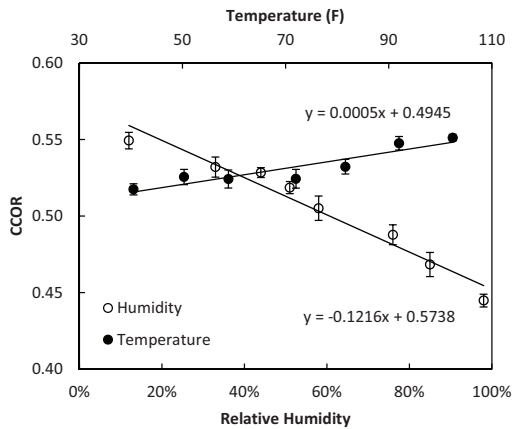} \caption{CCOR
as a function of relative humidity and temperature.} \label{fig:ccor}
\end{center}
\end{figure}

The CCOR is presented as a function of humidity level and temperature in
Fig.~\ref{fig:ccor}.
We find that the slope of the CCOR versus RH for MLB baseballs,  d(CCOR)/d(\%RH)=$-12.2\times 10^{-4}$, is considerably
larger than the flat-plate COR measured by Kagan and Atkinson\cite{kagan04} for NCAA baseballs, $-5.4\times
10^{-4}$.  The slope of the CCOR versus T, d(CCOR)/d(T)=$5\times 10^{-4}$/F, is about twice that measured for an NCAA baseball.\cite{drane04}
%Also new is the
%ball stiffness as a function of
%humidity and temperature shown in Fig.~\ref{fig:ds}.
%\begin{figure}[htb]
%\begin{center}
%\epsfig{angle=0,width=0.4\textwidth,file=fig10.eps} \caption{Ball
%stiffness as a function of relative humidity and temperature.} \label{fig:ds}
%\end{center}
%\end{figure}
%Ball stiffness is commonly reported as the compressive force to
%displace the ball 0.25 inches between flat platens (ASTM F1888).
%In the compression test the force increases with the contact area.
%Given two balls of the same material, but different diameters, the
%ball with the larger diameter will produce a higher compression.
%Comparison of softball and baseball compression values has led to
%the common misconception that softballs are harder than
%baseballs.\cite{ciocco08}  The ball stiffness test used here
%represents a more objective scale than compression to compare ball
%stiffness, independent of diameter.
%The effect of humidity on the
%stiffness was similar to the effect on CCOR.
%The error bars in Figs.~\ref{fig:w}-\ref{fig:ds} represent
%the standard deviation for each six-ball group. The softballs and
%baseballs have a comparable weight coefficient of variation (COV).
%The trend changes with CCOR and stiffness, however, where the
%baseball COV is twice that of the softballs. This difference in
%COV is surprising as the winding process used to make baseballs is
%usually considered to be more repeatable than the formulation and
%cure of polyurethane. The comparison suggests that the
%repeatability of baseballs may need as much scrutiny as softballs
%receive.

In view of the large difference between the present measurement of
the effect of RH on the CCOR for the MLB baseball and that found
by Kagan, it is worthwhile recalculating the effect of the humidor
on a long fly ball.  If the RH is increased from 30\% to 50\%, the
CCOR decreases by 0.024, or by about 4.5\% of its value.  We
assume a similar decrease occurs at the higher speeds of the
ball-bat impact.  For a typical MLB bat, pitch speed, and bat
speed, we estimate a decrease in batted ball speed by about 2.5
mph, corresponding to a decrease in fly ball distance by about 14.
ft.  Adair estimates that each percent change in fly ball distance
changes the probability of hitting a home run by about
7\%.\cite{adair02} Taking 380 ft. as a typical home run distance,
a reduction of 14 ft corresponds to a reduction in home run
probability by about 25\%, a very significant result and one that
is not inconsistent with the numbers quoted by Meyer and
Bohn.\cite{meyer08}  We conclude that it is plausible that the
humidor can account for the decrease in offensive statistics at
Coors Field since 2002.

A similar analysis can be done for the temperature dependence of the CCOR.
Balls stored at 70F will have a higher CCOR by 0.018 (or 3.3\%) than those stored at 35F.  Once again
assuming the same fractional change at higher speeds, a reduction of the temperature from 70F to
35F would lead to an decrease in fly ball distance
by about 10 ft, corresponding to an decrease in home run production by about 19\%.

\section{Summary}\label{sec:summary}
\ind We have addressed three practical questions at the
intersection of baseball and science.  We have shown that there is
no measurable trampoline effect with a corked bat and that it is
very unlikely that a batter can hit a baseball harder by using a
corked bat.  We have found no evidence that baseballs of today are
more or less lively than baseballs used in the late 1970's.
Finally, we have shown that storing baseballs in humidors at 50\%
relative humidity in Denver can lead to a marked reduction in
home run production.  A similar effect can be achieved by storing the baseballs at
a temperature of 35F.

\section*{Acknowledgments}
We gratefully acknowledge the Rawlings Corporation for supplying
the wood bats used in the corked bat study and to Nancy Finley for
providing the 1970's baseballs.

\end{document}